\documentclass[aip,jcp,reprint,showpacs,singlecolumn]{revtex4-1}

\usepackage{dcolumn}
\usepackage[normalem]{ulem}
\usepackage{placeins}
\usepackage{epsf}
\usepackage{subfigure}
\usepackage{epsfig,amsopn}
\usepackage{amssymb,graphicx}
\usepackage{amsmath,amsfonts}
\usepackage{graphicx}
\usepackage{bm}
\usepackage{ifpdf}
\usepackage{natbib}
\usepackage{wasysym}
\usepackage[byname]{smartref}
\usepackage{color}
\usepackage{lipsum}
\usepackage{nicefrac}
\usepackage{multirow}
\usepackage{braket}
\usepackage{float}
\usepackage{comment}
\usepackage{enumerate}
\usepackage{tabularx}
\usepackage[table]{xcolor}
\usepackage{xfrac}

\usepackage{amsmath,amsthm,verbatim,amssymb,amsfonts,amscd, graphicx}
\usepackage{braket}
\usepackage[breaklinks, colorlinks, linkcolor=blue, citecolor=blue, urlcolor=blue]{hyperref} 

\renewcommand{\v}[1]{\boldsymbol{#1}} 
\newcommand{\m}[1]{\boldsymbol{#1}} 

\newcommand{\beq}{\begin{equation}}
\newcommand{\eeq}{\end{equation}}

\newcommand{\bea}{\begin{eqnarray}}
\newcommand{\eea}{\end{eqnarray}}
\DeclareMathOperator{\Tr}{Tr}
\DeclareMathOperator{\cay}{cay}
\DeclareMathOperator{\Exp}{exp}

\begin{document}
	\title{Cayley modification for strongly stable path-integral and ring-polymer molecular dynamics}

\author{Roman Korol}
	\affiliation{Division of Chemistry and Chemical Engineering, California Institute of Technology, Pasadena, CA 91125, USA}
\author{Nawaf Bou-Rabee}
	\email{nawaf.bourabee@rutgers.edu}
	\affiliation{Department of Mathematical Sciences Rutgers University Camden, Camden, NJ 08102 USA}
\author{Thomas F. Miller III}
	\email{tfm@caltech.edu}
\affiliation{Division of Chemistry and Chemical Engineering, California Institute of Technology, Pasadena, CA 91125, USA}
\date{\today}
\begin{abstract}
Path-integral-based molecular dynamics (MD) simulations are widely used for the calculation of numerically exact quantum Boltzmann properties and approximate dynamical quantities.
A nearly universal feature of MD numerical integration schemes for equations of motion based on imaginary-time path integrals is the use of harmonic normal modes for the exact evolution of the free ring-polymer positions and momenta.
In this work, we demonstrate that this standard practice creates numerical artifacts.  In the context of conservative (i.e., microcanonical) equations of motion, it leads to numerical instability.  In the context of thermostatted (i.e., canonical) equations of motion, it leads to non-ergodicity of the sampling.  These pathologies are generally proven to arise at integration timesteps that depend only on the system temperature and the number of ring-polymer beads, and they are numerically demonstrated for the  cases of conventional ring-polymer molecular dynamics (RPMD) and thermostatted RPMD (TRPMD).
Furthermore, it is  demonstrated that these numerical artifacts are removed via replacement of the exact free ring-polymer evolution with a second-order approximation based on the Cayley transform.  The Cayley modification introduced here can immediately be employed with almost every existing integration scheme for path-integral-based molecular dynamics  --including path-integral MD (PIMD), RPMD, TRPMD, and centroid MD -- providing strong symplectic stability and ergodicity to the numerical integration, at no penalty in terms of computational cost, algorithmic complexity, or accuracy of the overall MD timestep.  Furthermore, it is shown that the improved numerical stability of the Cayley modification allows for the use of larger MD timesteps. We suspect that the Cayley modification will therefore find useful application in many future path-integral-based MD simulations.

\end{abstract}
	\maketitle


\section{Introduction}
\label{intro}

Feynman's path-integral formulation of quantum statistical mechanics \cite{Feynman1965} offers powerful and widely used strategies for including nuclear quantum effects in complex chemical systems. These strategies are based on the observation that the quantum Boltzmann statistical mechanics of a quantum system is exactly reproduced by the classical Boltzmann statistical mechanics of an isomorphic ring-polymer system.\cite{ChandlerWolynes} For the numerically exact calculation of quantum Boltzmann statistical properties, the classical Boltzmann distribution of the ring-polymer system can be sampled using   Monte Carlo\cite{ceperley1995revmodphys} (i.e., path-integral Monte Carlo, or PIMC) or molecular dynamics\cite{Parrinello1984} (PIMD).  For the approximate calculation of dynamical quantities, such as reaction rates,\cite{Craig2005,Craig2005a,CMDexamplerate} diffusion coefficients,\cite{Miller2005,Miller2005a,CMDexamplediffusion} and absorption spectra,\cite{Miller2005a,Habershon2008,Kaczmarek2009,Witt2009,CMDexamplespectrum} the Newtonian dynamics of the classical isomorphic system can be numerically integrated as a model for the real-time quantum dynamics, as in ring-polymer molecular dynamics (RPMD)\cite{Craig2004,Habershon2013} and centroid molecular dynamics (CMD).\cite{Cao1994,CMDReview}
These and related methods have enjoyed broad applicability in recent years for exploring nuclear quantum effects in the domains that span physical, bio-, geo-, and materials chemistry. \cite{Markland2018} 

For PIMD and RPMD calculations, considerable effort has been dedicated to the development and refinement of numerical integration schemes.
This work falls into two distinct categories.
In the first, the RPMD equations of motion are preconditioned by modifying the ring polymer mass
matrix; this causes the integrated trajectories to differ from those of the RPMD model,\cite{Martyna1999, Minary2003, BeRoStVo2008, BePiSaSt2011, Lu2018,Zhang2017, Liu2016} 
but it can lead to efficient and strongly stable\cite{BeRoStVo2008, BePiSaSt2011, Lu2018} 
sampling of the quantum Boltzmann distribution. 
In the second category, 
no modification is made to the ring-polymer mass matrix (i.e., the ``physical" masses of the ring-polymer beads are employed).\cite{Braams2006,Ceriotti2010,Ceriotti2011,Rossi2014,Rossi2018} 

Within the second category, it is  common to apply a thermostat to the internal ring-polymer motions, with two primary aims: to  more efficiently sample  the quantum Boltzmann distribution,\cite{Ceriotti2010,Ceriotti2011,Zhang2017} or to avoid the ``spurious resonance" artifact of the microcanonical (i.e., un-thermostatted) RPMD equations of motion in which internal ring-polymer modes mechanically couple to physical modes of the system.\cite{Rossi2014,Rossi2018}
PIMD and RPMD integration schemes in the second category (which preserve the RPMD model dynamics) typically employ a Trotter-like factorization of the time evolution operator.\cite{Tuckerman1993,Miller2005,Miller2005a,Ceriotti2010,Ceriotti2011,Rossi2014,Rossi2018}  For the example of thermostatted RPMD (TRPMD)\cite{Rossi2014}
using the generalized Langevin equation (GLE) thermostat,\cite{Ceriotti2010} the numerical integration is performed using\cite{Bussi2007} 
\begin{equation}
 e^{\Delta t L} 
=
e^{\frac{\Delta t}{2}L_\gamma}
e ^{\frac{\Delta t}{2}L_V}
e^{\Delta tL_0}
e^{\frac{\Delta t}{2}L_V}
e^{\frac{\Delta t}{2}L_\gamma}
+ \mathcal{O}(\Delta t^3)
	\label{eq:splitwithnoise}
\end{equation}
where the Liouvillian $L=L_V+L_0+L_\gamma$
includes contributions from the physical potential, $L_V$, the purely harmonic free ring-polymer motion, $L_0$, and the friction and thermal noise, $L_\gamma$;
note that the standard microcanonical RPMD numerical integration scheme is then recovered in the limit of zero coupling to the thermostat, such that \cite{Miller2005}
\begin{equation}
 e^{\Delta t L} 
=
e^{\frac{\Delta t}{2}L_V}
e^{\Delta t L_0}
e^{\frac{\Delta t}{2}L_V}
+ \mathcal{O}(\Delta t^3).
	\label{eq:split}
\end{equation}
Standard practice in these RPMD and PIMD integration schemes is to exactly evolve the harmonic free ring-polymer dynamics associated with $\exp(\Delta tL_0)$  using the uncoupled free ring-polymer normal modes.\cite{Tuckerman1993,Miller2005,Miller2005a,Ceriotti2010}

The first major conclusion of the current work is that any PIMD, RPMD, CMD,\cite{Stern2001,Poulsen2003,Ceriotti2016,Tuckerman1993,Craig2004,Habershon2013,Cao1994,Parrinello1984,Hone2006,Trenins2019,Benson2019,Ceriotti2010,Ceriotti2011,Rossi2014,Rossi2018} or other integration scheme that involves the exact integration of the free ring polymer (i.e., involves the ubiquitous $\exp(\Delta tL_0)$ step in terms of the ring-polymer normal modes) will exhibit provable numerical deficiencies, including resonance instabilities and non-ergodicity.
For the case of the standard microcanonical RPMD integration scheme in Eq.~\ref{eq:split}, which 
is a symplectic map, exact evolution of the  free ring-polymer step leads to the provable loss of strong symplectic stability and the demonstrable appearance of resonance instabilities in the integrated trajectories.
For thermostatted RPMD and PIMD integration schemes that involve a free ring-polymer  step,\cite{Ceriotti2010,Ceriotti2011,Rossi2014,Rossi2018} 
exact evolution of that step leads to the provable and numerically demonstrable non-ergodicity.

The second major conclusion of the current work is that these numerical artifacts can be eliminated by simply replacing the exact  evolution of the free ring polymer step with an approximation based on the Cayley transform: an alternative to exact free ring-polymer evolution that is no more costly, no more complicated, and no less accurate in the context of the full integration timestep.   
In particular, we show that this Cayley modification eliminates the resonance instabilities that occurs when trajectories are evolved using standard microcanonical RPMD integrators, and we show that it restores ergodicity to thermostatted RPMD and PIMD trajectories.  Furthermore, we show that the improved numerical properties of the Cayley modification generally allows for larger RPMD and PIMD integration timesteps to be employed.

The paper is organized as follows. In section~\ref{theory} we articulate the numerical instability problem in the context of standard RPMD numerical integration and introduce the Cayley modification as the solution. Section~\ref{nve} numerically illustrates the instability of standard RPMD numerical integration and shows that the Cayley modification removes this problem. Finally, in section~\ref{thermo} we generalize these findings to thermostatted trajectories.


\section{Theory}
\label{theory}

The theory introduced in this paper adapts and advances previous mathematical results on the numerical approximation of general second order Langevin stochastic partial differential equations with space-time white noise.\cite{Bo2017}

\subsection{RPMD}

We consider a quantum particle in 1D with Hamiltonian operator given by
\beq
	\hat{H}=\frac{\hat{p}^2}{2m}+V(\hat{q})
	\label{eq:H_q}
\eeq
where $\hat{q}$, $\hat{p}$, and $m$ represent the particle position, momentum, and mass, respectively, and $V(\hat{q})$ is a potential energy surface.  All results presented here are easily generalized to multiple dimensional quantum systems.

The thermal equilibrium properties of the system are described by the  quantum mechanical Boltzmann partition function,
\beq
	Q=\Tr[e^{-\beta \hat{H}}] \;, 
	\label{eq:Q_q}
\eeq
where $\beta=(k_B T)^{-1}$ is the inverse temperature. 
Using a path-integral discretization, $Q$ can be approximated by a classical partition function $Q_n$ of a ring-polymer with $n$ beads,\cite{Parrinello1984}
\beq
	Q_n=\frac{m^n}{(2\pi\hbar)^n}\int d^n\v{q}\int d^n\v{v}e^{-\beta H_n(\v{q},\v{v})} \;, 
	\label{eq:Q_n}
\eeq
where  $\v{q} = (q_0, \dots, q_{n-1})$ is the vector of bead positions, and $\v{v}$ is the corresponding vector of velocities. The ring-polymer Hamiltonian is given by
\beq
	H_n(\v{q},\v{v})=H_n^0(\v{q},\v{v})+
V^{\textrm{ext}}_n(\v{q}),
	\label{eq:H_n}
\eeq
which  includes contributions from the physical potential
\beq
V^{\textrm{ext}}_n(\v{q}) = \frac{1}{n} \sum_{j=0}^{n-1} V(q_j)
	\label{eq:Vext}
\eeq
and the free ring-polymer Hamiltonian
\beq
	H_n^0(\v{q},\v{v})=
	\frac{m_n}{2}\sum_{j=0}^{n-1}\left[v_{j}^2+
		\omega_n^2(q_{j+1}-q_{j})^2\right], 
	\label{eq:H_spring}
\eeq
where $m_n=m/n$, $\omega_n = n/(\hbar\beta)$ and $q_{n} = q_{0}$.

If we let $n=1$ in Eq.~\ref{eq:Q_n}, the classical partition function of the system (governed by a classical Hamiltonian, Eq.~\ref{eq:H_n} with $n=1$) is recovered, i.e. $Q_1=Q_{cl}$. 
In the limit $n\rightarrow\infty$, the path-integral approximation converges to the exact quantum Boltzmann statistics for the system, such that $Q_\infty=Q$. 
The thermal ensemble of ring-polymer configurations associated with Eq.~\ref{eq:Q_n} can be sampled using either molecular dynamics (leading to PIMD methods) or  Monte Carlo (leading to PIMC methods).

The classical equations of motion associated with the ring-polymer Hamiltonian in Eq.~\ref{eq:H_n},
\begin{eqnarray} \label{eq:rpmd_eoM_components}
 \dot{q}_j &=& v_j,\\
 \dot{v}_j &=& \omega_n^2 (q_{j+1}+q_{j-1} - 2 q_j) - \frac{1}{m} V'(q_j)\nonumber,
\end{eqnarray}
yield the RPMD model for the real-time dynamics of the system.\cite{Craig2004,Habershon2013}
RPMD provides a means of  approximately calculating Kubo-transformed thermal time-correlation functions, such as the position autocorrelation function \beq
	\tilde{C}_{qq}(t)=\frac{1}{Q}\Tr[e^{-\beta \hat{H}}\tilde{q}(0)\hat{q}(t)]
	\label{eq:C_Kubo}
\eeq
where the Kubo-transformed position operator $\tilde{q}$ is
\beq
	\tilde{q}=\frac{1}{\beta}\int_0^\beta e^{\lambda \hat{H}}\hat{q}e^{-\lambda\hat{H}}d\lambda
	\label{eq:Kubo}
\eeq
and the time-evolved operator $\hat{q}(t)$ is $e^{i\hat{H}t/\hbar}\hat{q}e^{-i\hat{H}t/\hbar}$. 

Specifically, the RPMD approximation to Eq.~\ref{eq:C_Kubo} is
\beq
	\tilde{C}_{qq}(t)=\frac{1}{Q_n}\int d^n\v{q}\int d^n\v{v}e^{-\beta H_n(\v{q},\v{v})} \bar{q}(0) \bar{q}(t)
	\label{eq:C_qq}
\eeq
where $\bar{q}$ is the bead-averaged position
\beq
	\bar{q}(t)=\frac{1}{n}\sum_{j=0}^{n-1} q_{j}(t) \;,  
	\label{eq:q_bead}
\eeq and the pair $(\v{q}(t), \v{v}(t))$ are evolved  by the RPMD equations of motion in Eq.~\ref{eq:rpmd_eoM_components} with initial conditions drawn from the classical Boltzmann-Gibbs measure.

 The RPMD equations of motion can  be compactly rewritten as \begin{equation} \label{eq:rpmd_eom}
\begin{bmatrix} 
\dot{\v{q}} \\ \dot{\v{v}}  
\end{bmatrix}
=  
\m{A}  
\begin{bmatrix} 
\v{q} \\ 
\v{v} 
\end{bmatrix}  
+  
\begin{bmatrix} 
\m{0} \\ 
\v{F}(\v{q})/m_n \end{bmatrix} 
\;,  \ \text{where} \ \m{A} = \begin{bmatrix} \m{0}  & \m{I} \\ \m{L} & \m{0} \end{bmatrix} \;,
\end{equation}
  $\v{F}( \v{q}) = -\nabla V^{\textrm{ext}}_n( \v{q} )$, $\m{I}$ is an $n \times n$ identity matrix, $\m{0}$ is an array of zeros, and
  $\m{L}$ is the $n \times n$ Toeplitz matrix 
 \beq
	\m{L} = \omega_n^2 \begin{bmatrix} 
	-2 & 1 & 0 & \cdots & 0 & 1 \\
1 & -2 & 1 & 0 & \cdots & 0 \\
 & \ddots & \ddots & \ddots \\
  && \ddots & \ddots & \ddots \\
0 & \cdots & 0 & 1 & -2 & 1 \\
1 & 0 & \cdots & 0 & 1 & -2
	\end{bmatrix}  \;.
	\label{eq:L}
\eeq
 We recognize $\m{L}$ as the 1D discrete Laplacian endowed with periodic boundary conditions; it is negative semi-definite with spectral radius that scales as $n^2$, and since $\v{L}$ is circulant, it can be diagonalized by the $n \times n$ real discrete Fourier transform (DFT) matrix.
In particular, the spectral decomposition of $\m{L}$ can be written as
\beq \label{eq:VDV}
\m{L} = - \m{U} \m{\Omega} \m{U}^{\mathrm{T}}, \quad  \text{where 
$\m{\Omega} = \operatorname{diag}(\omega_0^2, \dots, \omega_{n-1}^2)$} 
\eeq is a diagonal matrix of eigenvalues ordered in descending order and given by \begin{equation} \label{eq:eigenvalues}
\omega_j^2 = \begin{cases}
4 \omega_n^2 \sin^2\left( \frac{\pi j}{2 n} \right)
 & \text{if $j$ is even} \;, \\
4 \omega_n^2 \sin^2\left( \frac{\pi (j+1)}{2 n} \right)  & \text{else} \;, 
\end{cases}
\end{equation}
and $\m{U}$ is an $n \times n$ matrix whose columns are the corresponding orthonormal eigenvectors.

In nontrivial applications, the RPMD equations of motion in Eq.~\ref{eq:rpmd_eom} cannot be solved analytically.  It is then necessary to employ approximate numerical integration of the equations of motion.   
As we discuss next, designing good numerical integrators for Eq.~\ref{eq:rpmd_eom} is complicated by the interplay between the time-evolution of the free ring-polymer (obtained by setting $ \v{F}=0$ in  Eq.~\ref{eq:rpmd_eom}) and the contributions from the physical forces $\v{F}$.

\subsection{Cayley removes instabilities in a free ring-polymer mode}
\label{nve_1d}

RPMD is an example of highly oscillatory Hamiltonian dynamics. \cite{petzold_jay_yen_1997}   To understand why numerical integration of such systems
is tricky and why the Cayley modification is needed, it helps to consider the equations of motion for a particular normal mode of the free ring polymer with Matsubara frequency $\omega>0$: 
\begin{equation} \label{eq:lo}
\begin{bmatrix} \dot q \\ \dot v \end{bmatrix} = \m{A} \begin{bmatrix} q \\ v \end{bmatrix} \; \quad \text{where} \quad \m{A} = \begin{bmatrix} 0 & 1 \\ -\omega^2 & 0 \end{bmatrix}, 
\end{equation} which are also the equations of motion for a linear oscillator with natural frequency $\omega$.  If $\omega$ is large, Eq.~\ref{eq:lo} is highly oscillatory.   Solving Eq.~\ref{eq:lo} amounts to approximating the matrix exponential $\Exp(\Delta t \m{A})$ where $\Delta t$ is a timestep size. A good $2 \times 2$ matrix approximation $\m{M}_{\Delta t}$ should satisfy: \begin{description}
\item[(P1) Accuracy]  $\| \m{M}_{\Delta t} - \Exp(\Delta t \m{A}) \| =  O(\Delta t^3)$.
\item[(P2) Strong Stability]  For all $\omega >0$, and for all $\Delta t$ smaller than some constant independent of $\omega$, $\m{M}_{\Delta t}$ is a strongly stable symplectic matrix.
\item[(P3) Time-Reversibility] For all $\omega>0$ and $\Delta t>0$, $\m{M}_{\Delta t}$ is reversible with respect to the velocity flip matrix $\m{R} =\begin{bmatrix} 1 & 0 \\ 0 & -1 \end{bmatrix}$, i.e., $\m{R} \m{M}_{\Delta t} \m{R} = \m{M}_{\Delta t}^{-1}$.  
\end{description}

We briefly comment on each of these criteria for a good approximation.  Property (P1) is a basic requirement that ensures second-order accuracy on finite-time intervals. Property (P3) is particularly useful for sampling from the stationary distribution, since a reversible map can be readily Metropolized \cite{tierney1998note,BoVa2012,BoSaActaN2018}, and since time-reversibility in a volume-preserving numerical integrator  leads to a doubling of the accuracy order (see Propositions~5.2 and Theorem~6.2 of Ref.~\onlinecite{BoSaActaN2018}, respectively).  Property (P2) is the most interesting.  A symplectic matrix $\m{S}$ is
stable if all powers of the matrix $\m{S}$ are bounded.  A symplectic matrix $\m{S}$ is {\em strongly stable} if $\m{S}$ is stable and all sufficiently close symplectic matrices are also stable.  In other words, $\m{S}$ is strongly stable if there exists an $\epsilon>0$, such that all symplectic matrices $\m{S}^{\epsilon}$ that are within a distance $\epsilon$ away from $\m{S}$ are also stable.  A sufficient condition for $\m{S}$ to be strongly stable is that the eigenvalues of $\m{S}$ are on the unit circle in the complex plane and are distinct; both the necessary and sufficient conditions for strong stability of symplectic matrices are known.\cite{krein1950generalization}

\begin{figure}
\begin{center}
\includegraphics[width=0.44\textwidth]{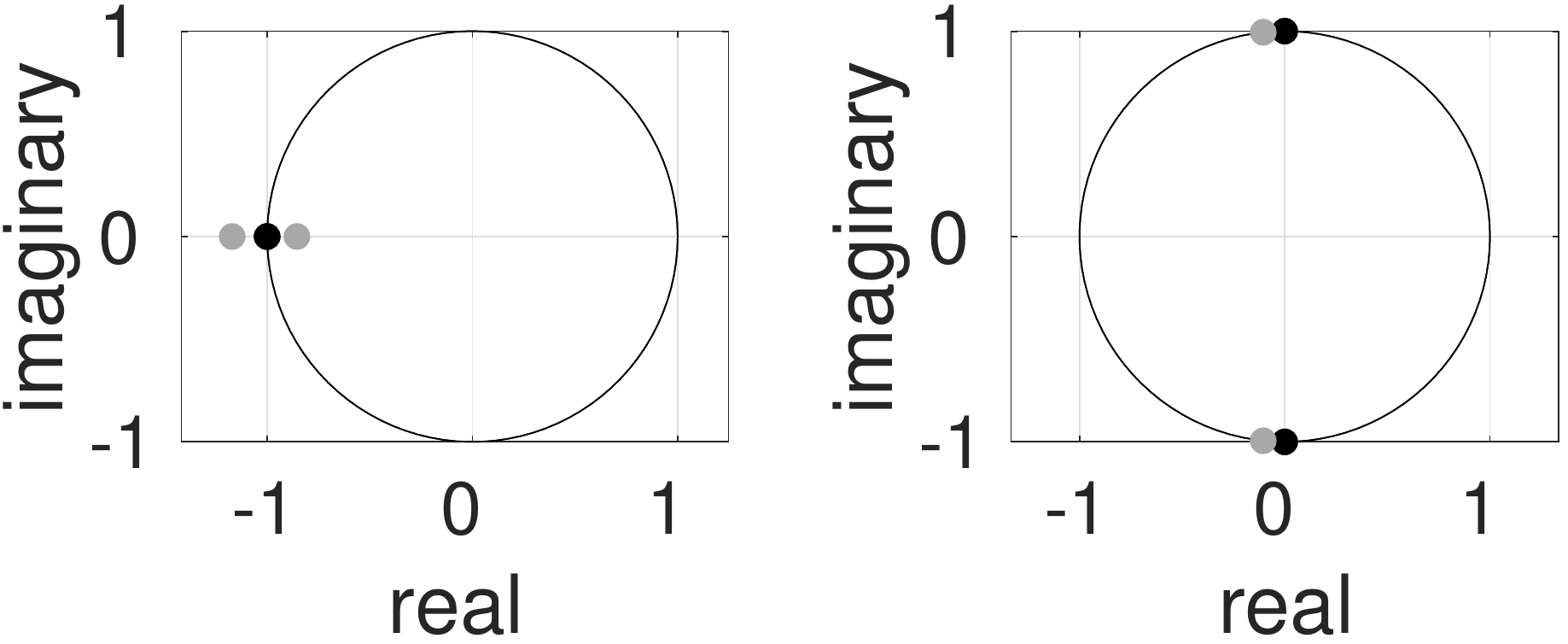}
\hbox{
   (a) $t=\frac{\pi}{3}$
        \hspace{1.0in} (b) $t=\frac{\pi}{4}$  
        } 
\end{center}
\caption{\small  {\bf Eigenvalues of $2\times2$ Symplectic Matrices.}
Eigenvalues of a symplectic matrix $\m{S} = \Exp(t \m{A})$  (black dots) are plotted in the complex plane along with eigenvalues of a perturbed symplectic matrix $\m{S}^{\epsilon} = \Exp((1/2) t \m{B}) \Exp(t \m{A}) \Exp((1/2) t \m{B})$  (grey dots).  The elements of $\m{A}$ and $\m{B}$ are specified in the text. 
For both values of $t$, $\m{S}$ is stable since its eigenvalues lie on the unit circle.  When the eigenvalues of $\m{S}$ are not distinct, then as shown in (a), $\m{S}^{\epsilon}$ has an eigenvalue with modulus greater than one, and hence, $\m{S}^{\epsilon}$ loses stability.  However, if the eigenvalues of $\m{S}$ are distinct, then $\m{S}$ is strongly stable, and as shown in (b), $\m{S}^{\epsilon}$ is stable since its eigenvalues remain on the unit circle.  
}
  \label{fig:strongly_stable}
\end{figure}

\begin{figure}
\begin{center}
\includegraphics[width=0.44\textwidth]{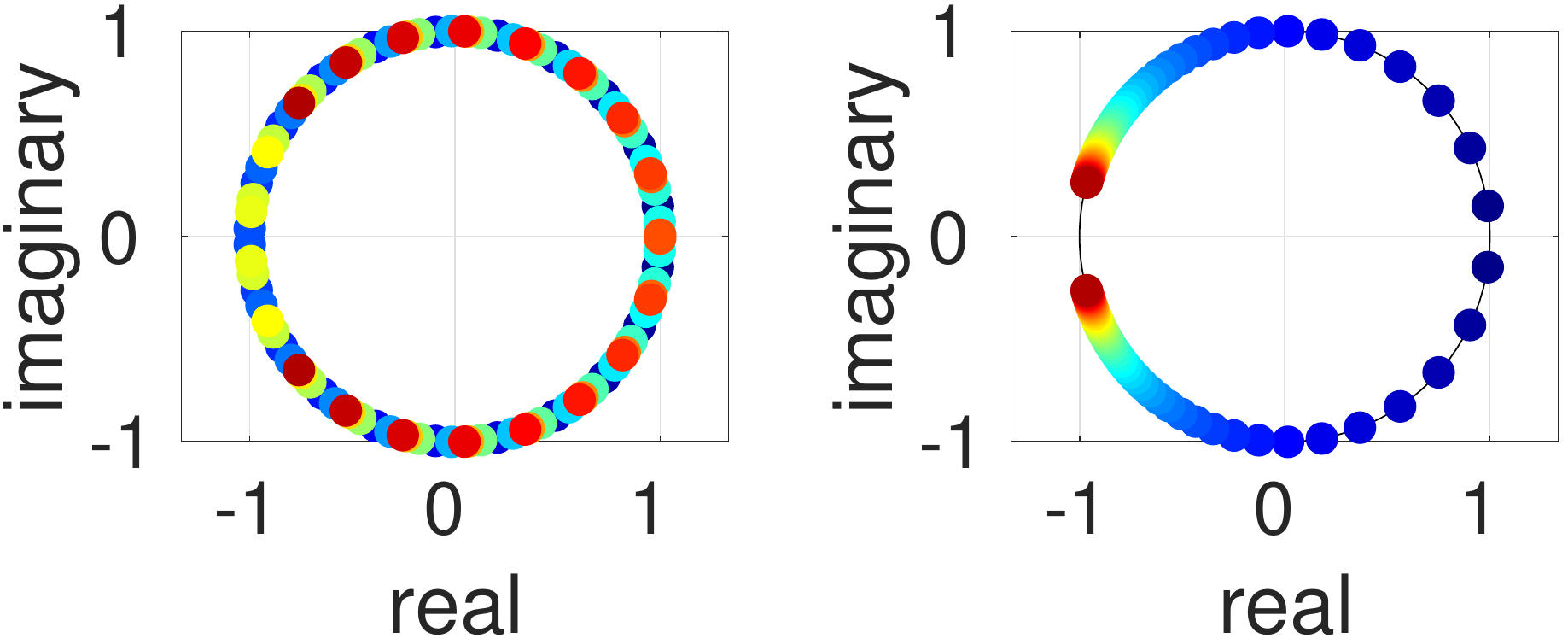}
\hbox{
   (a) $\exp(\Delta t \m{A})$
        \hspace{0.7in} (b) $\cay(\Delta t \m{A})$
        } 
\end{center}
\caption{\small  {\bf Eigenvalues of the Exponential vs.~Cayley Maps.}  
Eigenvalues of $\exp(\Delta t \m{A})$ (a) and $\cay( \Delta t \m{A})$ (b) at 
$50$ different timestep sizes between $0.05$ and $5.0$ (evenly spaced) and with $\omega=3$, color-coded from blue (smallest) through green and yellow to red (largest).  For $\exp(\Delta t \m{A})$, the eigenvalues rotate around
the unit circle multiple times.  However, for $\cay( \Delta t \m{A})$, the eigenvalues start near $(1,0)$, but never reach $(-1,0)$.  Since the
eigenvalues of $\cay( \Delta t \m{A})$ are always distinct, it provides strong symplectic stability,  whereas the matrix exponential loses strong stability every time the eigenvalues hit the horizontal axis.
In both panels, the eigenvalue associated with the ring-polymer centroid motion is excluded.
}
  \label{fig:eigenvalues_cayley_exp}
\end{figure}

Figure~\ref{fig:strongly_stable} illustrates the concept of strong stability.  In particular, for different values of $t$ (as indicated in each panel), the black dots correspond to the eigenvalues of the symplectic matrix $\m{S} = \Exp(t \m{A})$ with $\omega = 3$, and the grey dots are the eigenvalues of a perturbation of $\m{S}$ which preserves the symplectic nature of the matrix, specifically $\m{S}^{\epsilon} = \Exp((1/2) t \m{B}) \Exp(t \m{A}) \Exp((1/2) t \m{B})$ where $\m{B}=\begin{bmatrix} 0 & \epsilon \\ \epsilon & 0 \end{bmatrix}$ and $\epsilon=0.15$.  For any $t$, note that the two eigenvalues of $\m{S}$ are always on the unit circle, and hence, $\m{S}$ is always stable, but as the figure shows, $\m{S}$ is not always strongly stable.  Indeed, in Figure~\ref{fig:strongly_stable} (a), we see that the two eigenvalues of $\m{S}$, represented by a single black dot, are both equal to $(-1,0)$, which violates the condition for strong stability, and in this case, we see that one of the eigenvalues of $\m{S}^{\epsilon}$ has modulus greater than unity, which implies that $\m{S}^{\epsilon}$ is unstable.  In Figure~\ref{fig:strongly_stable} (b), the two eigenvalues of $\m{S}$ are distinct and equal to $(0,\pm1)$, and hence, $\m{S}$ is strongly stable. Since $\m{S}$ is strongly stable, and $\epsilon$ is sufficiently small, $\m{S}^{\epsilon}$ has eigenvalues that are on the unit circle, and hence, is itself stable. For a more detailed discussion of the concept of strong stability of symplectic matrices, see Section 42 of Ref.~\onlinecite{arnol2013mathematical}.

A natural candidate for an approximation $\m{M}_{\Delta t}$ that satisfies these criteria is the Verlet integrator, which is ubiquitous in the classical simulation of molecular systems.\cite{AlTi1987,FrSm2002,Bo2014,Leimkuhler2015}  For a single Matsubara frequency of the free ring polymer, the Verlet integrator gives \[
\m{M}_{\Delta t} = \begin{bmatrix} 1 - \frac{\Delta t^2 \omega^2}{2} & \Delta t \\ -\frac{1}{2} \Delta t \omega^2 (2 - \frac{\Delta t^2 \omega^2}{2}) & 1 - \frac{\Delta t^2 \omega^2}{2}  \end{bmatrix} \;.
\]
However, for $\Delta t > 2 / \omega$, the eigenvalues of $\m{M}_{\Delta t}$ are real and distinct, so that one of them has modulus $>1$, and therefore the powers of $\m{M}_{\Delta t}$ grow exponentially.
Thus, numerical stability requires $\Delta t < 2 / \omega$, and Verlet does not satisfy (P2), since this numerical stability requirement is not uniform in $\omega$.  

Surprisingly,  the exact solution for the normal-mode dynamics also does not satisfy (P2).  To see why, note that the eigenvalues of the matrix exponential $\Exp(\Delta t \m{A} )$ are $e^{ \pm i \omega \Delta t}$ and (P2) requires that  $e^{i \omega \Delta t} \ne e^{-i \omega \Delta t}$ which is violated if and only if \begin{equation} \label{eq:1d_strong_stability}
\Delta t = \frac{\pi k}{ \omega} \quad \text{for all $k \ge 1$} \;. 
\end{equation}  At these timesteps, the exact solution violates strong stability.  
This is illustrated in Figure~\ref{fig:eigenvalues_cayley_exp} (a), where the two eigenvalues of $\Exp(\Delta t \m{A})$ are plotted in the complex plane for a range of time-step sizes.  Although the two eigenvalues of $\Exp(\Delta t \m{A})$ lie on the unit circle for all $\Delta t$, strong stability fails to hold whenever the eigenvalues are both equal to $(\pm 1,0)$.

A simple strategy to avoid these artificial resonances is to use a random timestep size $\delta t$, e.g., take as timestep size an exponential random variable $\delta t$ with mean $\Delta t$.  Averaging $\Exp(\delta t \m{A})$ over the exponential probability density function yields $\m{M}_{\Delta t} = \mathbb{E}( \Exp(\delta t \m{A} ) ) = (\m{I} - \Delta t \m{A})^{-1}$, where here $\m{I}$ is the $2 \times 2$ identity matrix.  Unfortunately, as can be easily verified, this matrix satisfies none of our criteria: it is neither symplectic, nor reversible, nor sufficiently accurate.  However, we can easily turn this approximation into one that satisfies (P1), by simply composing $1/2$ step of this integrator with $1/2$ step of its adjoint $\m{M}_{\Delta t}^{-1}$.  This correction yields the Cayley transform of the matrix $\Delta t \m{A}$,
\begin{equation}
\cay(\Delta t \m{A}) \equiv (\m{I} - (1/2) \Delta t \m{A})^{-1} (\m{I} + (1/2) \Delta t \m{A}).\label{eq:cay}
\end{equation}

In fact, the Cayley transform satisfies all three of the specified criteria for a good numerical integrator. 
It is time-reversible since $\m{R} \cay(\Delta t \m{A}) \m{R} = (\m{R} - (1/2) \Delta t \m{R} \m{A})^{-1} (\m{R} + (1/2) \Delta t \m{A} \m{R}) = \cay(\Delta t \m{A})^{-1}$, where we used that $\m{R}^{-1} = \m{R}$.  It is a symplectic matrix since \[
\cay(\Delta t \m{A})^T \m{J} \cay(\Delta t \m{A}) = \m{J} \quad \text{where}  \quad  \m{J} = \begin{bmatrix} 0 & 1 \\ -1 & 0 \end{bmatrix} 
\] where we used the fact that $\m{A}$ is a Hamiltonian matrix (See Ref.~\onlinecite{MaRa1999}, Section 2.5).
More importantly, it is a strongly stable symplectic matrix for all $\Delta t>0$, as illustrated in Figure~\ref{fig:eigenvalues_cayley_exp} (b); in contrast with the exponential map,  for all $\omega>0$  and $\Delta t>0$ the  eigenvalues of the Cayley map are $(4 - \Delta t^2 \omega^2 \pm 4 i \Delta t \omega)  / (4+\Delta t^2 \omega^2 )$, which are distinct and of unit modulus.  Thus, not only is every matrix power of $\cay( \Delta t \m{A})$ bounded, but the Cayley map is strongly stable uniformly in $\omega$ and $\Delta t$.

\subsection{Cayley removes instabilities in microcanonical RPMD}

For numerical integration of the conservative RPMD equations of motion (Eq.~\ref{eq:rpmd_eoM_components} or Eq.~\ref{eq:rpmd_eom}), it is standard practice\cite{Miller2005, Miller2005a, Habershon2013} to employ a symmetrically split second-order integrator of the form in Eq.~\ref{eq:split}. 

Furthermore, it is standard practice to exactly perform the free ring-polymer time evolution step,\cite{Habershon2013}
using an exponential map of the form $\exp(\Delta tL_0)=\Exp(\Delta t \m{A})$ where $\m{A}$ is the matrix associated with the dynamics of the free ring-polymer Hamiltonian,
 \begin{equation} \label{eq:free_rpmd_eom}
\begin{bmatrix}   \dot{\v{q}} \\ \dot{\v{v}}  \end{bmatrix} =  
\m{A}  \begin{bmatrix} \v{q} \\ \v{v} \end{bmatrix}. \;
\end{equation}
In practice, the exact exponential map is executed by successively \emph{(i)} changing from the Cartesian bead positions and velocities to the normal modes of the free ring polymer, \emph{(ii)} numerically integrating each of the uncoupled normal mode equations of motion, and \emph{(iii)} translating the time-evolved normal mode coordinates back into the Cartesian bead positions and velocities. 
Therefore, the numerical stability of  standard RPMD numerical integration  may be analyzed in normal mode coordinates, where the free ring-polymer equations of motion in Eq.~\ref{eq:free_rpmd_eom} decouple into a system of $n$ independent oscillators with natural frequencies given by the eigenvalues of the matrix $\m{L}$ in Eq.~\ref{eq:eigenvalues}.

\begin{figure}
\begin{center}
\includegraphics[width=0.48\textwidth]{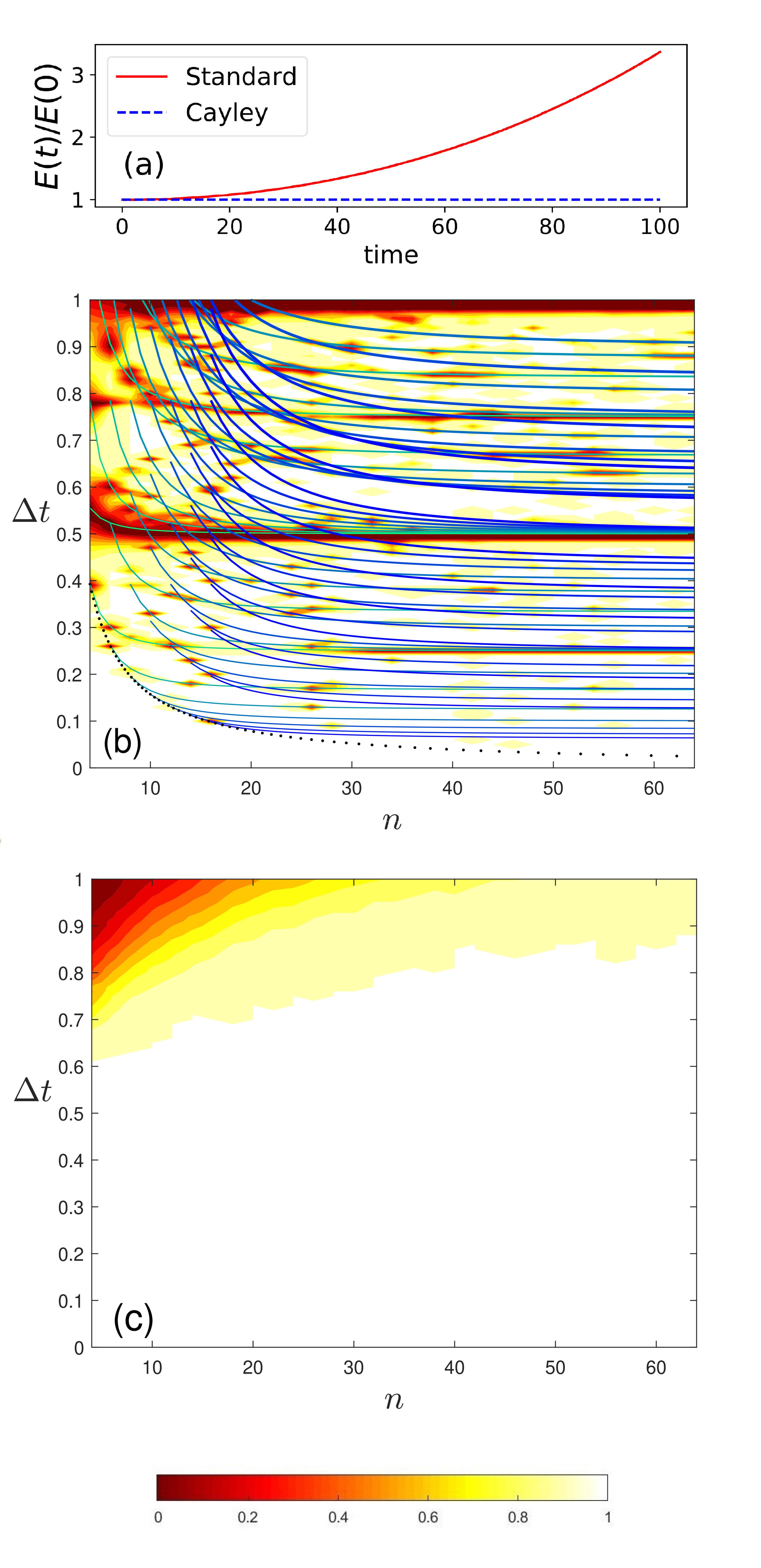}
\end{center}
\caption{\small {\bf Stability of RPMD Trajectories on the Harmonic Oscillator Potential.}
(a) Representative trajectories performed using the standard RPMD integration scheme and using the Cayley modification. 
(b) Results for the standard RPMD numerical integration.  The solid lines plot the instability condition in Eq.~\ref{eq:free_rp_stability} for $k=\{1,\dots,10\}$ and $j=\{2,4,\dots,16\}$.
Higher values of  $j$ are more blue, and higher values of $k$ are thicker.
The dotted black line shows the maximum safe timestep defined in Eq.~\ref{eq:max_safe_timestep}.
The heatmap indicates the fraction of stable  trajectories using  standard RPMD integration. 
(c) The the fraction of stable  trajectories using  Cayley-modified RPMD integration.
Results obtained at temperature $\beta=1$.
}
  \label{fig:free_rp_stability}
\end{figure}

By applying Eq.~\ref{eq:1d_strong_stability} to each normal mode coordinate, we find that strong stability of the exact free ring-polymer time evolution is violated when
\begin{equation}
    \label{eq:free_rp_stability}
\Delta t = \frac{\pi k}{\omega_j} \quad \text{for all $k \ge 1$ and $1 \le j \le n-1$} \;.
\end{equation}
Unstable pairs of $\Delta t$ and $n$ are plotted using solid lines in Fig.~\ref{fig:free_rp_stability}(b) for selected values of $j$ and $k$.  The horizontal asymptotes in this figure reflect the fact that the eigenvalues of $\m{L}$ converge to the eigenvalues of the continuous Laplacian endowed with periodic boundary conditions.

Unlike the exact free ring-polymer step used in standard RPMD numerical integration, the Cayley modification $\exp(\Delta tL_0)\approxeq \cay(\Delta t \m{A})$ is strongly stable for all $\Delta t>0$ uniformly in $n$.  To see this, note that the Cayley transform can be equivalently computed in either bead or normal mode coordinates.  More precisely, let $\m{L} = - \m{U} \m{\Omega} \m{U}^{\mathrm{T}}$ be the spectral decomposition of $\m{L}$ given in Eq.~\ref{eq:VDV}.  Direct computation then shows that \[
\cay(\Delta t \m{A} ) = \begin{bmatrix} \m{U} & \m{0} \\ \m{0} & \m{U} \end{bmatrix} 
\cay\left(\Delta t \begin{bmatrix} \m{0}  & \m{I} \\ - \m{\Omega} & \m{0} \end{bmatrix} \right)  \begin{bmatrix} \m{U}^{\mathrm{T}} & \m{0} \\ \m{0} & \m{U}^{\mathrm{T}} \end{bmatrix}  \;.
\] Using this correspondence, one can invoke the preceding results on the one-dimensional oscillator, to conclude that $\cay(\Delta t \m{A})$ is second-order accurate, strongly stable symplectic, and time-reversible.

Since the Cayley transform meets our criteria (P1)-(P3), and under suitable conditions on the force $\v{F}$, the Cayley modification to the RPMD numerical integrator is provably stable and second-order accurate on finite-time intervals with a stability requirement that is uniform with respect to the number of ring polymer beads.  On the other hand, standard RPMD integrators may display artificial resonance instabilities because the free RP step is not always strongly stable.  These instabilities often manifest as exponential growth in energy when strong stability is lost, as will be discussed in Section~\ref{nve}.

We emphasize that the improved numerical stability of the Cayley modification comes at zero cost in terms of algorithmic complexity or computational expense, and it preserves the same order of accuracy for the overall timestep.  Use of this improved integration algorithm simply involves replacing the exact normal mode free ring-polymer step in the standard RPMD integrator with the Cayley modification.

\subsection{Algorithmic comparison:  Standard vs.~Cayley}
\label{alg_comp}
For complete clarity, we now present a side-by-side comparison of the full RPMD timestep
(Eq.~\ref{eq:split}) with the free ring-polymer motion $\exp(\Delta t L_0)$ implemented using either the standard exponential map (i.e., exact normal mode evolution) or via the Cayley modification.  In both cases, the full RPMD timestep associated with the splitting in Eq.~\ref{eq:split} is implemented using the algorithm
\begin{equation} \label{RPMDtimestep}
\begin{array}{lc}
\textbf{Velocity half-step:}   &  \v{v} \leftarrow \v{v} + \frac{\Delta t}{2} \frac{\v{F}}{m_n}\\
\textbf{Free ring-polymer step:\quad}   &     (\v{q},\v{v}) \leftarrow \textrm{FRP}(\v{q},\v{v}; \Delta t) \\
\textbf{Force evaluation:}    &    \v{F} = - \nabla V^{\textrm{ext}}_n(\v{q})\\
\textbf{Velocity half-step:}    &     \v{v} \leftarrow \v{v} + \frac{\Delta t}{2} \frac{\v{F}}{m_n}
\end{array}
\end{equation}

In standard RPMD numerical integration, the free ring-polymer step is performed exactly, using:
\begin{enumerate}
    \item Convert bead Cartesian coordinates to normal modes using the orthogonal transformation:
	\beq \label{eq:nm_transform}
	\v{\varrho} =  \m{U} \v{q} \qquad \text{and} \qquad 
    	\v{\varphi} =  \m{U} \v{v}
	\eeq
	where $\m{U}$ is the real DFT matrix defined in Eq.~\ref{eq:VDV}.
    \item From $t$ to $t + \Delta t$, exactly evolve the free ring polymer in the normal mode coordinates:
\beq
	\begin{pmatrix} \varrho_j(t+\Delta t) \\ \varphi_j(t+\Delta t)\end{pmatrix} = \Exp(\Delta t \m{A}_j)
	\begin{pmatrix} \varrho_j(t) \\ \varphi_j(t)\end{pmatrix}
	\label{eq:frp_step}
\eeq
where \[
\m{A}_j = \begin{bmatrix} 0 & 1 \\
-\omega_j^2 & 0 \end{bmatrix} \;, 
\]
for $0 \le j \le n-1$ with $\omega_j$ defined in Eq.~\ref{eq:eigenvalues}.
    \item Convert back to bead Cartesian coordinates using the inverse of $\m{U}$, which is just its transpose, since $\m{U}$ is orthogonal.
\end{enumerate}

In the Cayley modification, the only change is to use the following in place of Eq.~\ref{eq:frp_step}: 
\beq
	\begin{pmatrix}\varrho_j(t+\Delta t) \\  \varphi_j(t+\Delta t)\end{pmatrix} = \cay(\Delta t \m{A}_j)
	\begin{pmatrix} \varrho_j(t) \\ \varphi_j(t)\end{pmatrix},  
	\label{eq:cay_evolve}
\eeq
where $\cay$ is the Cayley transform given in Eq.~\ref{eq:cay}.

As a final algorithmic comparison, we note that  another popular means of evolving the free ring-polymer involves   multiple timestepping (MTS) with the reversible reference system propagator algorithm (RESPA) \cite{Tuckerman1992,Martyna1999,Tuckerman1993}, which introduces an inner loop of short  timesteps.  However,  it is  easily shown that  MTS-RESPA  can exhibit the same problem of resonance instabilities as exact normal-mode evolution, due to fact that MTS-RESPA also lacks the property of strong stability.  Consequently, we will not further discuss MTS algorithms in the current work, although we recognize that combining the Cayley modification with MTS in the context of the ring-polymer contraction method\cite{Markland2008,Markland2008b} is straightforward and worth pursuing.


\section{Results for RPMD}
\label{nve}
In this section, we demonstrate the numerical integration of the microcanonical RPMD equations of motions (Eq.~\ref{eq:rpmd_eom}).  Specifically, we compare the performance of the standard RPMD integrator, which involves exact integration of the free ring-polymer modes (Eq.~\ref{eq:frp_step}) and our refinement in which the Cayley modification is used (Eq.~\ref{eq:cay_evolve}).  
Results are presented for simple one-dimensional potentials, including
\begin{align}
  \text{Harmonic:}~ V(q)&=\frac{1}{2}q^2     \label{eq:harm}\\
\text{Weakly anharmonic:} ~ V(q)&=\frac{1}{2}q^2+\frac{1}{10}q^3+\frac{1}{100}q^4  \label{eq:anharm}\\
\text{Quartic:} ~ V(q)&=\frac{1}{4}q^4  \label{eq:quart}
\end{align}
and using a mass of $m=1$.

We begin by numerically testing the conditions for loss of strong stability (Eq.~\ref{eq:free_rp_stability})
for the example of the harmonic potential (Eq.~\ref{eq:harm}). Figure~\ref{fig:free_rp_stability}(a) shows a typical example of one of the approximately $25\%$ of trajectories that fail for the standard RPMD integration scheme with $\beta=1$, $n=16$, and $\Delta t=0.1$.
 The unstable trajectories start out with the typical values of ring-polymer energy in Eq.~\ref{eq:H_n} (i.e. they are not the ``hot" initial conditions from the tail of the thermal distribution), and they diverge to exponentially large energies after relatively short propagation time when run with the standard RPMD. All of these trajectories are stable when run with the Cayley modification.
 
The solid lines in Fig.~\ref{fig:free_rp_stability}(b) indicate predicted conditions for instability (Eq.~\ref{eq:free_rp_stability}).  These analytical predictions are overlaid with a heatmap showing the fraction of stable RPMD trajectories on the harmonic potential using the standard RPMD integration scheme; for the purposes of the current section, a trajectories is deemed to be unstable if
energy conservation associated with the ring-polymer Hamiltonian (Eq.~\ref{eq:H_n}) is violated by more than $10\%$ within 100 time units of simulation. 
There are clear correlations in Fig.~\ref{fig:free_rp_stability}(b) between the predicted instabilities and observed simulation results.

Finally, Fig.~\ref{fig:free_rp_stability}(c) presents the corresponding heatmap for the Cayley-modified RPMD integration scheme. 
The Cayley modification preserves the conditions for strong stability, and the only numerically unstable trajectories are found for extremely large timesteps ($\Delta t>0.6$).
Comparison of Figs.~\ref{fig:free_rp_stability}(b) and (c) reveals the clear numerical advantages of the Cayley-modified RPMD integration scheme over the standard RPMD integration scheme. 

Before proceeding, we emphasize the generality of the loss of strong stability with the standard RPMD numerical integrator:  Eq.~\ref{eq:free_rp_stability} makes no assumption with regard to the form of the physical potential, the dimensionality of the system, or the mass of the particles; it  only depends on the temperature of the system and the number of ring-polymer beads in relation to the size of the integration timestep.  
Considering Eq.~\ref{eq:free_rp_stability} for the $k=1$ index and the highest Matsubara frequency of the ring-polymer, it is straightforward to show that the smallest possible timestep $\Delta t_{*}$ at which strong stability is violated is given by
\begin{equation}
    \label{eq:max_safe_timestep}
\Delta t_{*} = \frac{\beta\hbar\pi}{2n} \;.
\end{equation}
We thus arrive at a highly practical expression for the ``maximum safe timestep"  that depends only on $\beta$ and $n$, such that all smaller timesteps avoid the loss of strong stability associated with Eq.~\ref{eq:free_rp_stability}. 
In Fig.~\ref{fig:free_rp_stability}(b), this result is plotted (dotted, black line) and seen to follow the convex hull of smallest timesteps created by the other curves.
In passing, we note that if  $\beta$ corresponds to room temperature and $n=64$, then the maximum safe timestep is $0.6$ fs, which is strikingly consistent with the conventional $0.5$ fs timestep employed in many PIMD simulations of liquid water.

\begin{figure}
\begin{center}
\includegraphics[width=0.48\textwidth]{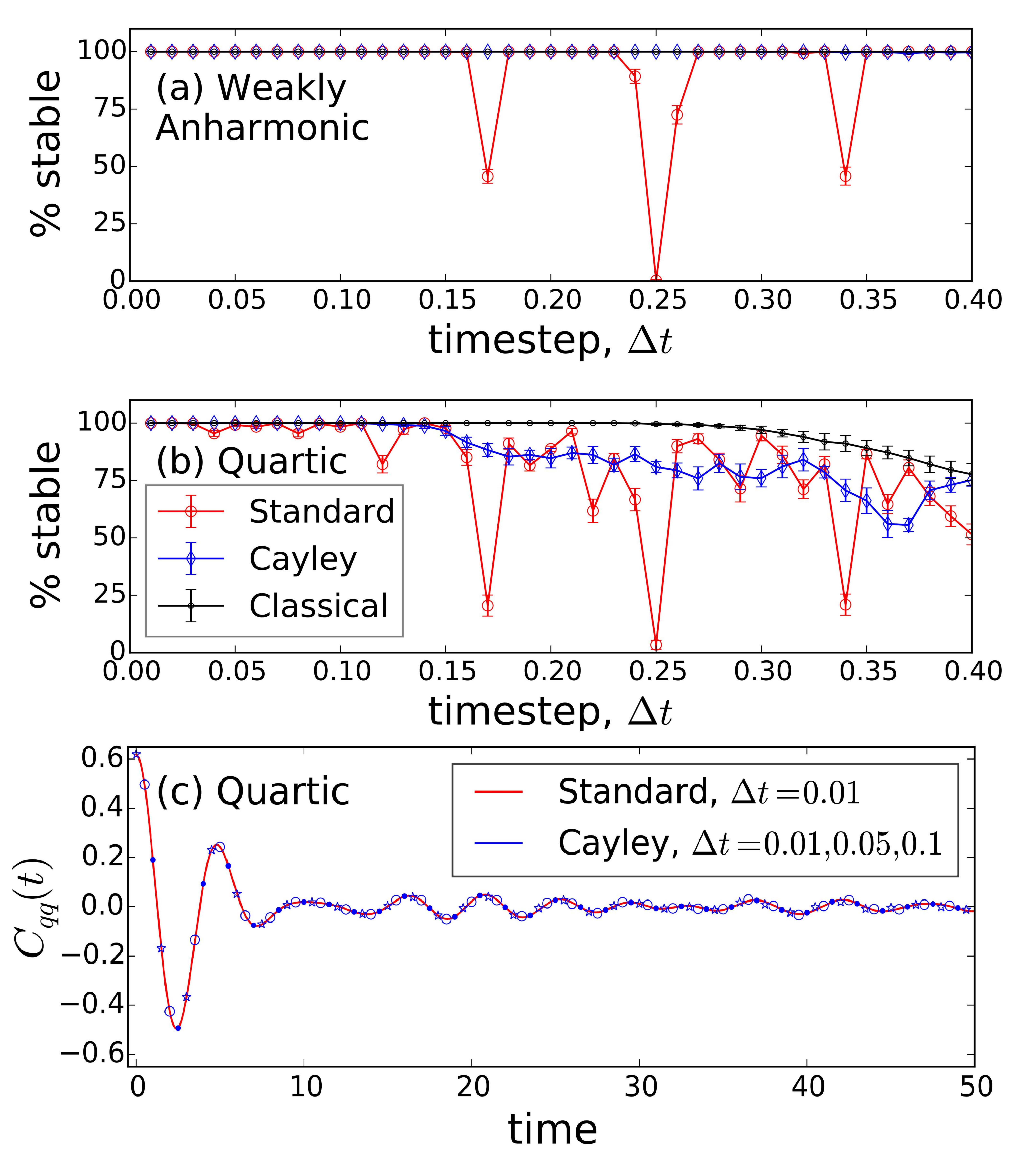}
\end{center}
\caption{\small  
{\bf Stability and Accuracy of RPMD Trajectories on Anharmonic Potentials.}
Percentage of stable RPMD trajectories using standard and Cayley-modified  integration as a function of timestep, for the (a)  weakly anharmonic and (b) quartic potentials.  Results obtained using $n=54$ and $\beta=1$.  Also included are classical MD results using the Verlet integrator. (c) For the quartic potential, comparison of the RPMD position time autocorrelation function obtained using standard integration with a small time-step where it is stable ($\Delta t = 0.01$) and using the Cayley modifiction with a range of larger timesteps ($\Delta t = 0.01$, filled circles; $\Delta t = 0.05$, empty circles; $\Delta= 0.10,$ stars), indicating no significant loss of accuracy.
}
\label{fig:percent_stable}
 
\end{figure}

Figure~\ref{fig:percent_stable} confirms that the numerical instabilities of the standard RPMD integrator also manifest for anharmonic potentials.  For both the weakly anharmonic (Eq.~\ref{eq:anharm}) and  quartic (Eq.~\ref{eq:quart}) potentials, we plot the fraction of stable trajectories as a function of timestep, comparing the standard RPMD integration scheme with the Cayley modification.  Also shown are the fraction of stable classical mechanical trajectories (i.e., the 1-bead limit of RPMD) when integrated using the Verlet algorithm.  Indeed, the standard RPMD integration scheme exhibits clear numerical instabilities at particular timesteps (which depend on the choice of $\beta$ and $n$), whereas the Cayley-modified integration scheme (like the classical integration scheme) avoids these pronounced instabilities.

For the results in Fig.~\ref{fig:percent_stable}, the maximum safe timestep is $\Delta t_{*} \approx 0.029.$ Note that the standard RPMD integration scheme on the weakly anharmonic potential does not exhibit significant loss of stability at this timestep, due to the fact that the unstable ring-polymer mode apparently does not sufficiently couple to the other modes on the timescale of the trajectories. However, the expected artifact at this timestep is indeed observed for the quartic potential.  These results illustrate that the degree to which the resonance instabilities of  standard RPMD integration manifest will depend on the application, but regardless of the system, these resonance instabilities can be removed using the Cayley modification.
Finally, panel (c) in this figure compares the accuracy of the standard and Cayley-modified RPMD integration schemes for the case of the quartic oscillator, revealing that even with time-steps that three-fold exceed the maximum safe timestep of the standard integration scheme, the Cayley-modified scheme shows negligible loss of accuracy in the trajectories.

\begin{figure}
\begin{center}
\includegraphics[width=0.48\textwidth]{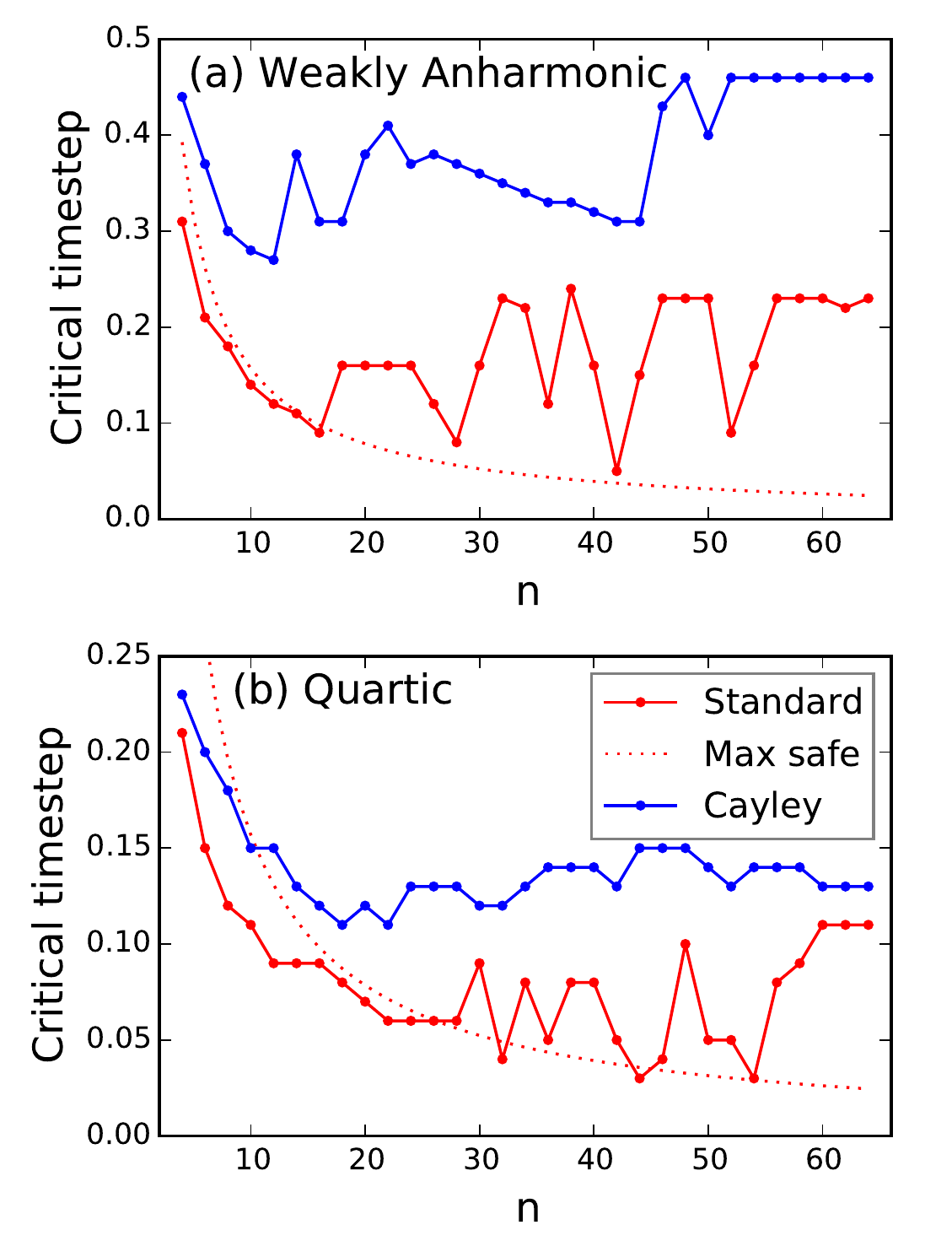}
\end{center}
\caption{\small {\bf Comparing largest stable timestep} as a function of the number of ring-polymer beads for the standard and Cayley-modified RPMD integration schemes on the (a) weakly anharmonic and (b) quartic potentials. The critical timestep for the numerical simulations is defined in the text. 
Also shown is the maximum safe timestep for the standard RPMD integration scheme (red dots). 
For classical MD integration using the Verlet algorithm, the critical timestep is  $0.5$ for the weakly anharmonic potential and $0.3$ for the quartic potential.
Results obtained at temperature $\beta=1$.}
  \label{fig:critical_timestep}
\end{figure}

Figure~\ref{fig:critical_timestep} explores the degree to which the Cayley modification enables the use of larger timesteps in comparison to the standard RPMD integration scheme.  Defining the ``critical timestep'' as the largest value of $\Delta t$ for which 980 out of 1000 trajectories are stable, we compare this quantity for  standard and Cayley-modified RPMD numerical integration as a function of the number of ring-polymer beads; the trends in the figure are insensitive to the precise definition of the critical timestep.
Also shown is the maximum safe timestep for the standard RPMD integration scheme (Eq.~\ref{eq:max_safe_timestep}).
The improved stability of the Cayley-modified integration scheme is seen to consistently allow for the use of larger RPMD timesteps. 
The numerical behavior of the standard RPMD integration scheme closely tracks the predictions of the maximum safe timestep, although as seen previously, the resonance instabilities do not always manifest on the timescale of the simulated trajectories.  Interestingly, for small $n$ in the quartic-oscillator simulations, the standard RPMD integration scheme actually underperforms the prediction of the maximum safe timestep, given that it exhibits large energy fluctuations ($>10\%$) without fully encountering a resonance instability.
In summary, using the maximum safe timestep for the standard RPMD integration scheme as a reference, the figure indicates that in these systems, the Cayley modification allows for substantial improvements in the allowed timestep size (three-fold or more for large $n$).

\section{Results for TRPMD}
\label{thermo}

Thermostatted RPMD (TRPMD) involves  thermalization of the internal ring-polymer modes during RPMD dynamics, with the aims of improving sampling of the Boltzmann distribution\cite{Ceriotti2010} or avoiding the ``spurious resonance" artifact that can appear in RPMD simulations of vibrational spectra.\cite{Habershon2008,Rossi2014} 
Following Refs.~\onlinecite{Ceriotti2010} and \onlinecite{Rossi2014}, we implement TRPMD using the splitting in Eq.~\ref{eq:splitwithnoise},
where $L_T$ corresponds to
\beq
    \dot{\v{v}}=-\m{\gamma}\v{v}+\sqrt{2n m^{-1} \beta^{-1}}\m{\gamma}^{1/2}\dot{\v{W}}(t),
    \label{eq:lang}
\eeq
 $\dot{\v{W}}(t)$ is a white-noise vector (since $\v{W}$ is an $n$-dimensional standard Brownian motion), and $\v{\gamma}$ is an $n \times n$ friction matrix defined such that $\m{U}^{\mathrm{T}} \v{\gamma} \m{U}$ is a diagonal matrix whose $k$th diagonal entry is equal to $\omega_k$ (Eq.~\ref{eq:eigenvalues}).  In normal mode coordinates (cf.~Eq.~\ref{eq:nm_transform}), this thermostat is implemented  by adding the following at the beginning and end of the full integration step outlined in Eq.~\ref{RPMDtimestep}:
\[
\varphi_j(t+\Delta t) = e^{- \frac{\omega_j \Delta t}{2}} \varphi_j(t) + \sqrt{n m^{-1} \beta^{-1}} \sqrt{1 - e^{-\omega_j \Delta t}} \xi_j \;,
\]
where $\xi_j$ is a standard normal variate.

\subsection{Cayley removes non-ergodicity in TRPMD}

Given that it helps to avoid spurious resonances,\cite{Rossi2014,Rossi2018} one might expect that a Langevin thermostat can also eliminate the instabilities we have observed in standard RPMD integrators.  This turns out to be only partly true.  Here, we show that \emph{(i)} lack of strong stability in the free RP step induces non-ergodicity in standard TRPMD integrators, and \emph{(ii)} the Cayley modification eliminates these non-ergodicity issues.

For this purpose, we revisit the simple case of a single free ring-polymer mode, as in Section~\ref{nve_1d}. Consider Eq.~\ref{eq:lo} with a Langevin thermostat, \begin{equation} \label{eq:trpmd_eoM_1d}
\begin{bmatrix} \dot q \\ \dot v \end{bmatrix} = \m{K} \begin{bmatrix} q \\ v \end{bmatrix} + \begin{bmatrix} 0 \\ \sqrt{2 \beta^{-1} \gamma} \dot W \end{bmatrix} \;, ~ \m{K} = \m{A} + \begin{bmatrix} 0 & 0 \\ 0 & -\gamma \end{bmatrix} \;,
\end{equation}
where $\gamma \ge 0$ is a friction factor and $\dot W(t)$ is a scalar white noise.   The solution $(q(t), v(t))$ of Eq.~\ref{eq:trpmd_eoM_1d} is a bivariate Gaussian with mean vector and covariance matrix given respectively by \begin{equation} \label{eq:law_langevin_1D}
\begin{aligned}
\v{\mu}(t) &= \Exp(t \m{K}) \begin{bmatrix} q(0) \\ v(0) \end{bmatrix} \;, \\ 
\m{\Sigma}(t) &= 
2 \beta^{-1} \gamma   \int_0^t \Exp(s \m{K})  \begin{bmatrix} 0 & 0 \\ 0 & 1 \end{bmatrix} \Exp(s \m{K}^{\mathrm{T}}) ds \;.
\end{aligned}
\end{equation}
In the limit as $t \to \infty$, the probability distribution of $(q(t), v(t))$ converges to the classical Boltzmann-Gibbs measure, which in this case, is a bivariate normal distribution with mean vector and covariance matrix given respectively by \begin{equation} \label{eq:equilibruM_law_langevin_1D}
\v{\mu} =   \begin{bmatrix} 0 \\ 0 \end{bmatrix} \;, \quad  \m{\Sigma} = \beta^{-1} \begin{bmatrix} \omega^{-2} & 0 \\ 0 & 1 \end{bmatrix} \;.
\end{equation}

In this situation, the standard TRPMD splitting in Eq.~\ref{eq:splitwithnoise} inputs $(q_0, v_0)$ and outputs $(q_1, v_1)$ defined as  \begin{equation} \label{eq:trpmd_schemes_1D} \begin{bmatrix} q_1 \\ v_1 \end{bmatrix}  = \m{O} \m{E}  \m{O} \begin{bmatrix} q_0 \\ v_0 \end{bmatrix} +  \sqrt{\frac{1-e^{-\gamma \Delta t}}{\beta}} \left(  \m{O} \m{E} \begin{bmatrix} 0 \\  1 \end{bmatrix} \xi_0 +   \begin{bmatrix} 0 \\  1 \end{bmatrix} \eta_0 \right)
\end{equation} where $\xi_0, \eta_0$ are independent standard normal random variables, $\m{E} = \Exp( \Delta t \m{A})$, and $\m{O}$ is the $2 \times 2$ matrix  \[
 \m{O} = \Exp\left(  \frac{\Delta t}{2} \m{\Gamma} \right) \;, \quad  \m{\Gamma} = \begin{bmatrix} 0 & 0 \\ 0 & -\gamma \end{bmatrix}  \;.
\]
Moreover,  the numerical solution after $N$ integration steps is a Gaussian vector with mean vector and covariance matrix given respectively by  \begin{equation} \label{eq:law_ltrpmd_schemes_1D}
\v{\mu}_N =  (\m{O} \m{E}  \m{O})^N \begin{bmatrix} q_0 \\ v_0 \end{bmatrix} \;, ~ 
\m{\Sigma}_N = \sum_{j=0}^{N-1}  (\m{O} \m{E}  \m{O})^j \m{Q}  (\m{O} \m{E}^{\mathrm{T}}  \m{O})^j \;,
\end{equation}
where  \[
 \m{Q} =\beta^{-1} (  1-e^{-\gamma \Delta t} )  \left( \m{O} \m{E}  \begin{bmatrix} 0 & 0 \\ 0 & 1 \end{bmatrix} \m{E}^{\mathrm{T}} \m{O} +   \begin{bmatrix} 0 & 0 \\ 0 & 1 \end{bmatrix}  \right) \;.
\]  From Eq.~\ref{eq:1d_strong_stability}, if $\Delta t = k \pi/\omega$ for any $k \ge 1$, then $\m{E}$ is not strongly stable.  At these timesteps, the eigenvalues of the matrix $\m{O} \m{E}  \m{O}$ are given by 
$\lambda_+ = (-1)^k$ and $\lambda_- = (-1)^k \exp(-k \pi \gamma/\omega)$. 
By the Cayley-Hamilton theorem for $2 \times 2$ matrices,\cite{Andreescu:2016:ELA} we have the following representation of the $N$th power of $\m{O} \m{E}  \m{O}$ 
\begin{align*}
& (\m{O} \m{E}  \m{O})^N =  \frac{(\lambda_+)^N}{\lambda_+ - \lambda_-} ( \m{O} \m{E}  \m{O} - \lambda_- \m{I} )  \\
& \quad +  \frac{(\lambda_-)^N}{\lambda_- - \lambda_+} ( \m{O} \m{E}  \m{O} - \lambda_+ \m{I} )  \;.
\end{align*} Since $|\lambda_+| = 1$, it follows from this representation that  $\v{\mu}_N$ does not converge to $\v{\mu}$ in Eq.~\ref{eq:equilibruM_law_langevin_1D}, since  $\v{\mu}_N$ clearly depends on the initial condition.  Similarly, the covariance matrix $\m{\Sigma}_N$ fails to converge to $\m{\Sigma}$.

If we modify the above by replacing every instance of $\m{E}$ with $\m{C} = \cay(\Delta t \m{A})$, the modified splitting is ergodic.  More precisely, provided that the timestep is sufficiently small such that
\begin{equation} \label{eq:tcay_stability_1d}
2 > (1 + \cosh(\gamma \Delta t) ) \left( \frac{4 - \Delta t^2 \omega^2}{4 + \Delta t^2 \omega^2} \right)^2,
\end{equation} then the eigenvalues of $\m{O} \m{C}  \m{O}$ are a complex conjugate pair 
with complex modulus  $|\lambda_{\pm}| = \exp(-\gamma \Delta t/2)$.  Hence, the matrix $\m{O} \m{C}  \m{O}$ is asymptotically stable.  Under condition \ref{eq:tcay_stability_1d}, the Cayley-modified scheme converges to the exact classical Boltzmann-Gibbs measure, in this example.

These results carry over to TRPMD, where the free ring-polymer equations of motion in Eq.~\ref{eq:free_rpmd_eom} decouple into a system of $n$ independent oscillators with natural frequencies given by the eigenvalues of the matrix $\m{L}$ in Eq.~\ref{eq:eigenvalues}. 
Although the  analysis of TRPMD in this section was performed for the specific case of the splitting in Eq.~\ref{eq:splitwithnoise} (i.e., the Bussi-Parrinello or OBABO splitting), we have confirmed that the same problem of non-ergodicity arises in the BAOAB splitting\cite{Leimkuhler2013} and can likewise be fixed via the Cayley modification.

\subsection{TRPMD  numerical results}

\begin{figure}
\begin{center}
\includegraphics[width=0.45\textwidth]{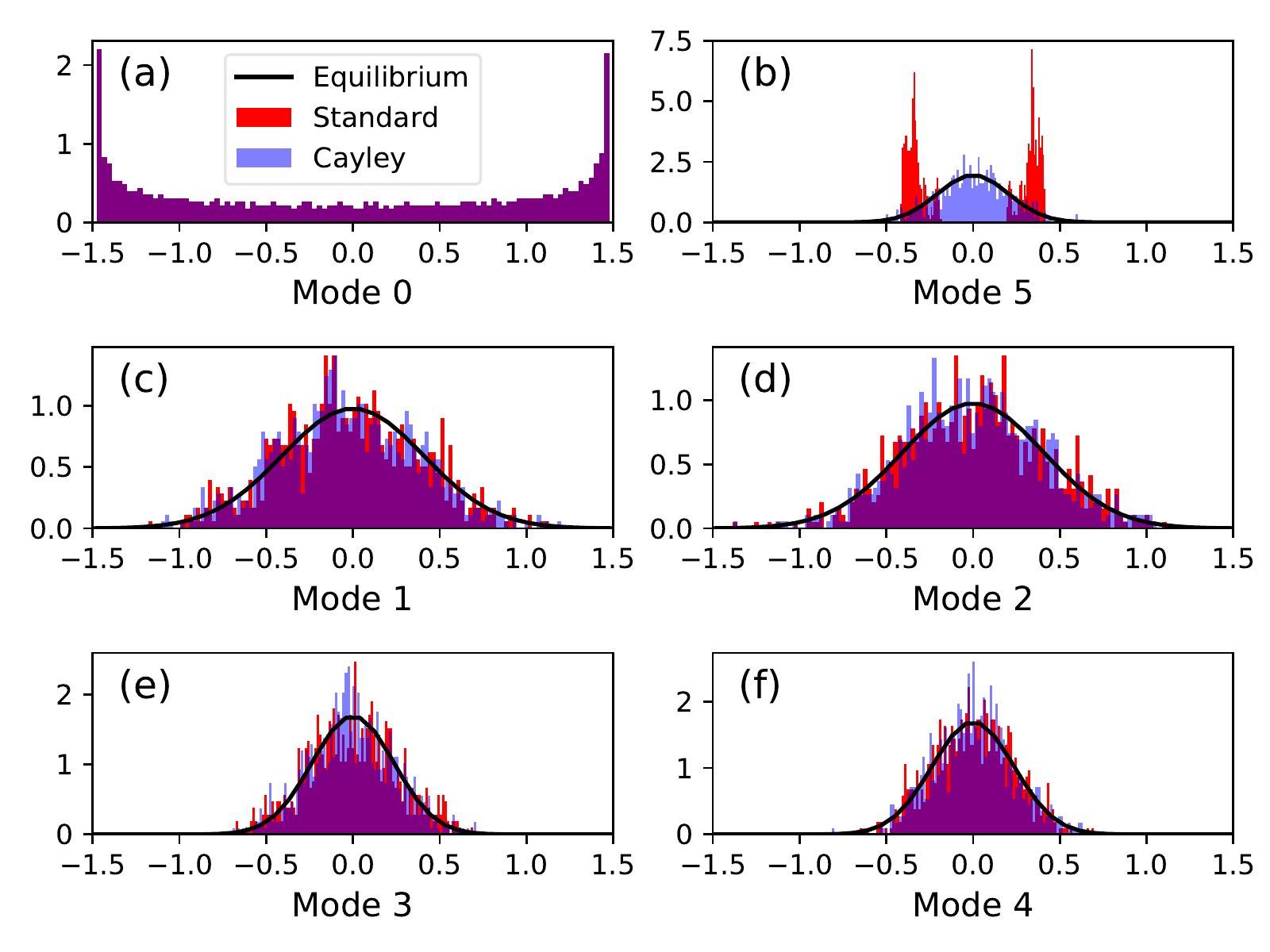}
\end{center}
\caption{\small  {\bf Ergodicity of TRPMD recovered with the Cayley modification, Example 1.} Normalized histograms of the ring-polymer normal mode displacement coordinates for a single trajectory (6 beads, $\beta=1$), evolved on the harmonic potential with a timestep of $\Delta t=0.26$. (a) The centroid mode, $\omega_j=0$. (b) The predicted non-ergodic mode with $\omega_5=12$, (c-d), (e-f) pairs of modes with $\omega_1=\omega_2=6$ and $\omega_3=\omega_4=10.4$, respectively. Solid black line indicate the equilibrium distribution of the internal modes. 
}
  \label{fig:nonerg}
\end{figure}

Figure~\ref{fig:nonerg}  presents TRPMD results on the harmonic potential (Eq.~\ref{eq:harm}) using $n=6$ and $\beta=1$.  For a single TRPMD trajectory, we histogram the distribution of the normal mode coordinates that are sampled, employing the smallest timestep for which numerical instability is observed in the microcanonical case for this number of beads (see Fig.~\ref{fig:free_rp_stability}b); specifically, we use $\Delta t=0.26$, which corresponds to the instability condition in Eq.~\ref{eq:free_rp_stability} for the case of $n=6$, $j=5$, and $k=1$.
Using both  standard and Cayley-modified TRPMD integration, the trajectory is sampled at every timestep for a total of 770 timesteps.

The centroid mode (panel a) follows harmonic motion, that is decoupled from the other degrees of freedom. With both integrators, the lower-frequency ($j=1-4$) internal ring-polymer modes are efficiently sampled and converge to the correct Gaussian distribution (panels c-f).
However, the $j=5$ mode behaves qualitatively differently, as predicted by Eq.~\ref{eq:free_rp_stability}, with the standard TRPMD integrator showing clear non-ergodicity.  The Cayley modification leads to ergodic sampling of all ring-polymer modes.

\begin{figure}
\begin{center}
\includegraphics[width=0.45\textwidth]{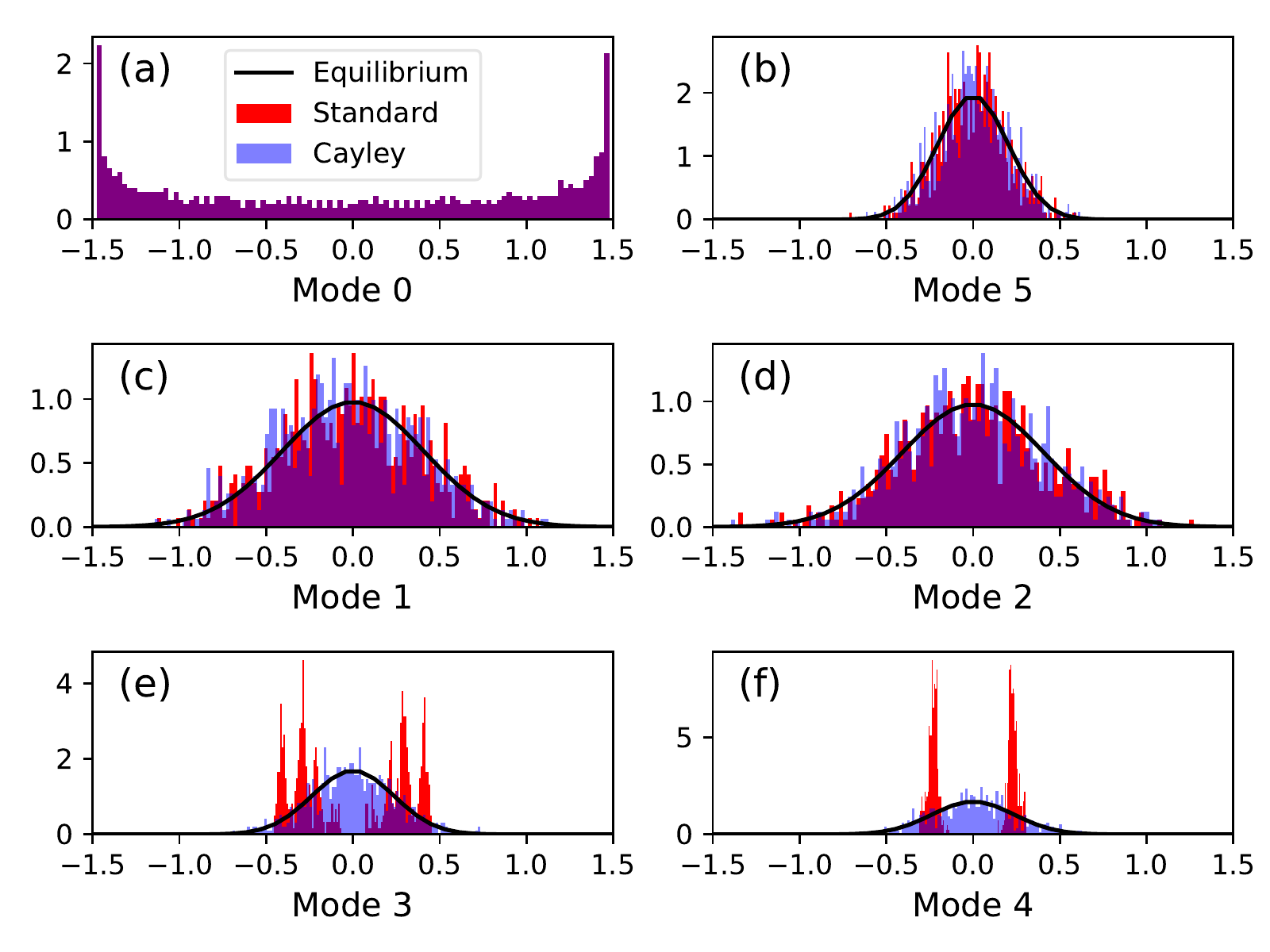}
\end{center}
\caption{\small  {\bf Ergodicity of TRPMD recovered with the Cayley modification, Example 2.} Normalized histograms of the ring-polymer normal mode displacement coordinates for a single trajectory (6 beads, $\beta=1$), evolved on the harmonic potential with a timestep of $\Delta t=0.3$. (a) The centroid mode, $\omega_j=0$. (b) Unique highest frequency mode with $\omega_5=12$, (c-d) Modes with $\omega_1=\omega_2=6$ (e-f) The predicted non-ergodic modes, $\omega_3=\omega_4=10.4$. Solid black line indicate the equilibrium distribution of the internal modes.
}
  \label{fig:nonerg2}
\end{figure}

The lower frequency internal modes can also be afflicted with non-ergodicity at larger timesteps in this system. For the next-smallest unstable timestep in Fig.~\ref{fig:free_rp_stability} ($\Delta t=0.3$, which corresponds to the instability condition in Eq.~\ref{eq:free_rp_stability} with $j=3,4$, and $k=1$), the simulations were repeated. 
As predicted by the instability condition,  modes 3 and 4 are found to be non-ergodic if sampled using the standard TRPMD integrator (Fig.~\ref{fig:nonerg2}); again, ergodicity is recovered using the Cayley modification.
The same non-ergodicity problems appear for anharmonic potentials using the standard TRPMD integrator and can  easily  be avoided with use of the Cayley modification.

We emphasize that the TRPMD results presented here employ a white-noise thermostat; there are additional non-ergodicity problems for coloured-noise TPRMD which the Cayley modification is not expected to improve, since they likely arise from the attenuation of the thermostat across particular frequency bands.\cite{Ceriotti2011,Ceriotti2013}

\section{Summary}

Strong stability is a relevant -- and  under-appreciated -- concept for path-integral-based molecular dynamics methods.
Without strong stability, numerical integration schemes are prone to numerical instabilities in the microcanonical case and non-ergodicity in the canonical case.  Fortunately, one can easily imbue existing integration schemes, including those for PIMD, RPMD, TRPMD, and many CMD methods, with strongly stability via the Cayley modification introduced here.
This can be done without downside in terms of the computational cost, algorithmic complexity, or accuracy of the numerical integration scheme.
The numerical results presented here suggest that this will have practical benefits for simulation studies, including improved stability, improved sampling efficiency, and improved efficiency via the use of larger MD timesteps.

While the Cayley transformation is familiar in the chemical physics literature in the context of the Crank-Nicolson propagator\cite{Crank1947} for wavepacket dynamics,\cite{Recepies,Judson1991} and real-time path integrals\cite{Ma1993} it has not to our knowledge been utilized  for molecular dynamics, due to an under-appreciation of the property of strong stability.
We conclude by noting that path-integral-based MD methods are far from unique in the physical sciences in exhibiting highly oscillatory  dynamics, with other notable examples including Markov-Chain-Monte-Carlo-based Bayesian statistical inversion,\cite{kaipio2005statistical,dashti2017bayesian,borggaard2018bayesian} transition path sampling, \cite{ReVa2005,pinski2010transition,Bolhuis2002,Miller2007} stochastic wave equations,\cite{NeVa2016}
Drude-oscillator models for many-body  polarizability and dispersion,\cite{DrudeOscillatorsMartyna,TkatchenkoMBDispersion,Jordancite} 
and Carr-Parrinello molecular dynamics.\cite{CPMD} 
 We anticipate that the Cayley modification introduced here may have similar advantages in these and other areas of application.

\label{sum}
\begin{acknowledgements}
We thank Jes\'{u}s Sanz-Serna, Xuecheng Tao, and Eric Vanden-Eijnden for helpful discussions.
N.~B.-R.~was supported in part by the National Science Foundation under Award No.~DMS-1816378.
R.~K.~and T.~F.~M.~acknowledge support from the Department of Energy under Award No.~DE-FOA-0001912 and the Office of Naval Research under Award No.~N00014-10-1-0884.
\end{acknowledgements}
\bibliography{RPMD}

\end{document}